\documentclass[12pt,aps,prb,preprint]{revtex4}
\usepackage{amsmath}
\usepackage{amsfonts}
\usepackage{amssymb}
\usepackage{graphicx}

\begin{document}
\title{Mapping of shape invariant potentials by the point canonical transformation }
\author{Mohammad R. Setare$^{1}$\thanks{E-mail:rezakord@ipm.ir},
 Ebrahim
Karimi,$^{2}$\thanks{E-mail:e\_karimi@uok.ac.ir}\\
 $^1${\small
Department of Science,  Payame Noor University. Bijar. Iran.} \\
$^2${\small \small Dipartimento di Scienze Fisiche,
Universit$\grave{a}$ di Napoli "Federico II", Complesso di Monte}\\
{\it \small  S. Angelo, via Cintia, 80126 Napoli, ITALY}\\}
\begin{abstract}
In this paper by using the method of point canonical transformation
we find that the Coulomb and Kratzer potentials can be mapped to the
Morse potential. Then we show that the P\"{o}schl-Teller potential
type I belongs to the same subclass of shape invariant potentials as
Hulth\'{e}n potential. Also we show  that the shape-invariant
algebra for Coulomb, Kratzer, and  Morse potentials is $SU(1,1)$,
while the shape-invariant algebra for P\"{o}schl-Teller type I and
Hulth\'{e}n is $SU(2)$.
\end{abstract}

\maketitle
\section{Introduction}
 It is known that all analytically solvable
potentials in quantum mechanics have the property of shape
invariance~\cite{gend}. In fact shape invariance is an integrability
condition, however, one should emphasize that shape invariance is
not the most general integrability condition as not all exactly
solvable potentials seem to be shape invariance
to~\cite{cooper,dabro}. An interesting feature of supersymmetric
quantum mechanics is that for a shape invariant system
\cite{cooper1, infeld} the entire spectrum can be determined
algebraically without ever referring to underlying differential
equations.\\
In this paper we briefly describe supersymmetric quantum mechanics,
then by using the method of point canonical transformation we find
that the Coulomb and Kratzer potentials can be mapped to the Morse
potential \cite{cooper2}.  The Kratzer potential \cite{kratzer} we
consider in this paper has played an important role in the history
of the molecular and quantum chemistry and it has been so far
extensively used to describe the molecular structure and
interactions \cite{mol}. After that we show that the
P\"{o}schl-Teller potential type I belongs to the same subclass of
shape invariant potentials as Hulth\'{e}n potential.
 The Hulth\'{e}n potential
\cite{h1,h2} is one of the important short-range potentials in
physics.  This potential is a special case of the Eckart potential
\cite{eckart} which has been widely used in several branches of
physics and its bound-state and scattering properties have been
investigated by a variety of techniques.
\section{Supersymmetry Quantum Mechanics and Shape invariance}
According to the factorization method~\cite{{Infeld},{dong}}, the
quantum mechanical Hamiltonian, after subtracting the ground energy,
is written as the product of an operator $\hat{A}$ and its Hermitian
conjugate, $\hat{A}^{\dag}$
\begin{equation}\label{1}
    \hat{H}-E_{0}=\hat{A}^{\dag}\hat{A}
\end{equation}
where $E_{0}$ is the ground state energy, and $\hat{A}$,
$\hat{A}^{\dag}$ are given by
\begin{eqnarray}\label{2}
    \hat{A}=W(x)+\frac{i}{\sqrt{2\,m}}\hat{p}\\\label{3}
    \hat{A}^{\dag}=W(x)-\frac{i}{\sqrt{2\,m}}\hat{p}
\end{eqnarray}
By definition (\ref{1}) the ground state wave function satisfies the
following condition
\begin{equation}\label{4}
    \hat{A}|\psi_{0}>=0
\end{equation}
Since the ground-state wave function $\psi_{0}(x)$ for a bound state
has no node, it can be written as
\begin{equation}\label{5}
    \psi_{0}(x)=e^{-\frac{\sqrt{2\,m}}{\hbar} \int{W(x)\,dx}}
\end{equation}
Using (\ref{4})  we have the following supersymmetric partner
Hamiltonian
\begin{equation}\label{6}
    \hat{H}_{1}=\hat{A}^{\dag}\hat{A},
    \hspace{15mm}\hat{H}_{2}=\hat{A}\hat{A}^{\dag}
\end{equation}
The corresponding potentials are given as
\begin{equation}\label{7}
    V_{1}=W^{2}(x)-\frac{\hbar}{\sqrt{2\,m}}\frac{d\,W(x)}{dx}
\end{equation}
\begin{equation}\label{8}
    V_{2}=W^{2}(x)+\frac{\hbar}{\sqrt{2\,m}}\frac{d\,W(x)}{dx}
\end{equation}
The Hamiltonian in (\ref{1}) is called shape-invariant\cite{dabro}
if the following condition is satisfied:
\begin{eqnarray}\label{9}
    \hat{A}(a_{1})\hat{A}^{\dag}(a_{1})=\hat{A}^{\dag}(a_{2})\hat{A}(a_{2})+R(a_{a_{1}})
\end{eqnarray}
where $a_{1}$, and $a_{2}$ represent the parameters of the
Hamiltonian. One can rewrite the above condition in term of the
partner potentials as:
\begin{eqnarray}\label{10}
    V_{2}(x,a_{1})=V_{1}(x,a_{2})+R(a_{1})
\end{eqnarray}
shape-invariant problem was formulated in algebraic terms
in\cite{Balan}. We assume that replacing $a_{1}$ by $a_{2}$ in a
given operator can be achieved with a similarity transformation
\begin{equation}\label{11}
    \hat{T}(a_{1}){\cal{O}}(a_{1})\hat{T}^{-1}(a_{1})={\cal O}(a_{2})
\end{equation}
There are two classes of shape-invariant potentials. For the first
class the parameters $a_{1}$ and $a_{2}$ of the two suppersymmetric
parameters are related to each other by
translation~\cite{cooper,Chuan}
\begin{equation}\label{12}
    a_{2}=a_{1}+\eta
\end{equation}
For the second class, the parameters $a_{1}$ and $a_{2}$ are related
to each other by scaling~\cite{khare1,Barc}
\begin{equation}\label{13}
    a_{2}=q\,a_{1}
\end{equation}
For the first class the operator $\hat{T}(a_{1})$ of (\ref{11}) is
given by
\begin{equation}\label{14}
    \hat{T}(a_{1})=e^{\eta\frac{\partial}{\partial a_{1}}},
    \hspace{15mm}\hat{T}^{-1}(a_{1})=\hat{T}^{\dag}(a_{1})
\end{equation}
In the second class, the similarity transformation (\ref{11}) is
given by following operator
\begin{equation}\label{15}
    \hat{S}(a_{1})=e^{\ln{q}\,a_{1}\frac{\partial}{\partial a_{1}}},
    \hspace{15mm}\hat{S}^{-1}(a_{1})=\hat{S}^{\dag}(a_{1})
\end{equation}
By introducing new operators
\begin{equation}\label{16}
      \hat{B}_{+}=\hat{A}^{\dag}(a_{1})\hat{T}(a_{1}),
      \hspace{15mm}\hat{B}_{-}=\hat{B}^{\dag}_{+}=\hat{T}^{\dag}(a_{1})\hat{A}(a_{1})
\end{equation}
the Hamiltonian can be rewritten as
\begin{equation}\label{17}
    \hat{H}-E_{0}=\hat{A}^{\dag}\hat{A}=\hat{B}_{+}\hat{B}_{-}
\end{equation}
Using Eqs.(\ref{9}) and (\ref{16}), one can obtain  following
commutation relation
\begin{equation}\label{18}
    [\hat{B}_{-},\hat{B}_{+}]=R(a_{0})
\end{equation}
where
\begin{equation}\label{19}
   a_{n}=a_{0}+n\,\eta\hspace{15mm}or\hspace{15mm}a_{n}=q^{n}\, a_{0}
\end{equation}
also following identities
\begin{equation}\label{20}
    R(a_{n})=\hat{T}(a_{1})R(a_{n-1})\hat{T}^{\dag}(a_{1}),\hspace{15mm}
    R(a_{n})=\hat{S}(a_{1})R(a_{n-1})\hat{S}^{\dag}(a_{1})
\end{equation}
valid for any $n$. By using Eqs.(\ref{16},\ref{20}) we can establish
the commutation relations
\begin{eqnarray}\label{21}
    [\hat{H},\hat{B}^{n}_{+}]&=&(R(a_{1})+R(a_{2})+\ldots+R(a_{n}))\hat{B}^{n}_{+}\\\label{22}
    [\hat{H},\hat{B}^{n}_{-}]&=&-\hat{B}^{n}_{-}(R(a_{1})+R(a_{2})+\ldots+R(a_{n}))
\end{eqnarray}
means that, $B^{n}_{+}|\psi_{0}>$ is an eigenstate of the
Hamiltonian with the eigenvalue $R(a_{1})+R(a_{2})+\ldots+R(a_{n})$.
The normalized eigenstate is
\begin{equation}\label{23}
    |\psi_{n}>=\frac{1}{\sqrt{R(a_{1})+\ldots+R(a_{n})}}\hat{B}_{+}\times
    \ldots\times\frac{1}{\sqrt{R(a_{1})+R(a_{2})}}\hat{B}_{+}\times\frac{1}
    {\sqrt{R(a_{1})}}\hat{B}_{+}|\psi_{0}>
\end{equation}
In addition to the oscillatorlike commutation relations
Eq.~(\ref{21}) one gets the commutation relations
\begin{equation}\label{24-1}
    [\hat{B}_{+},R(a_{0})]=\{R(a_{1})-R(a_{0})\}\hat{B}_{+}
\end{equation}
\begin{equation}\label{24-2}
    [\hat{B}_{+},[\hat{B}_{+},R(a_{0})]]=(\{R(a_{2})-R(a_{1})\}-\{R(a_{1})-R(a_{0})\})
    \hat{B}^2_{+}
\end{equation}
and so on.

\section{Mapping of Kratzer and Coulomb potentials to the Morse  potential}
Consider the following potential (We are using units with $\hbar=1$,
$2\,m=1$.)
\begin{equation}\label{24}
    V(x)=-\frac{\alpha}{x}+\frac{\beta}{x^2}+\gamma
\end{equation}
If we take $\alpha=\beta=1$ and $\gamma=0$, we obtain the Kratzer
potential as
\begin{equation}\label{25}
    V(x)=-(\frac{1}{x}-\frac{1}{x^2})
\end{equation}
in another case we take $\alpha=e^2$, $\beta=l(l+1)$ and
$\gamma=\frac{e^4}{4(l+1)^2}$, in this case the potential (\ref{24})
is as
\begin{equation}\label{26}
    V(x)=-\frac{e^2}{x}+\frac{l(l+1)}{x^2}+\frac{e^4}{4(l+1)^2}
\end{equation}
which is equivalent with Coulomb potential in 3-Dimension
Schr\"{o}dinger equation in spherical coordinates.\\
At first, we briefly review the method of mapping of shape-invariant
under point canonical transformation. For given potential of
Eq.~(\ref{24}), one can write the Schr\"{o}dinger equation as
\begin{equation}\label{5-1}
    \{-\frac{d^2}{dx^2}+V(\alpha_{i};x)-E(\alpha_{i})\}\psi(\alpha_{i};x)=0
\end{equation}
here $\alpha_{i}$ are the set of parameters of given potential
Eq.~(\ref{24}). Under a point canonical transformation, as following
\begin{equation}\label{5-2}
    x:=f(z),\hspace{10mm}\psi(\alpha_{i},x):=g(z)\,\widetilde{\psi}(\widetilde{\alpha}_{i};z)
\end{equation}
the Schr\"{o}dinger equation (\ref{5-1}) is transformed into
\begin{equation}\label{5-3}
    \{-\frac{d^2}{dz^2}+(\frac{f''}{f'}-2\,\frac{g'}{g})\frac{d}{dz}+(\frac{g'}{g}\,\frac{f''}{f'}-\frac{g''}{g})
    f'^{2}(V(\alpha_{i};f(z))-E(\alpha_{i}))\}\widetilde{\psi}(\widetilde{\alpha}_{i};z)=0
\end{equation}
or in the familiar form as
\begin{equation}\label{5-4}
    \{-\frac{d^2}{dz^2}+\widetilde{V}(\widetilde{\alpha}_{i};z)-\widetilde{E}(\widetilde{\alpha}_{i})\}
    \widetilde{\psi}(\widetilde{\alpha}_{i};z)=0
\end{equation}
in which $\alpha_{i}$ represent set of parameters of the transformed
potential,and the prime denotes differential with respect to the
variable $z$. To remove the first-derivative term from
Eq.~(\ref{5-3}), one requires
\begin{equation}\label{5-5}
    g(z)=C\sqrt{f'(z)}
\end{equation}
Using Eq.~(\ref{5-5}) and comparing Eqs.~(\ref{5-2},\ref{5-3}) we
obtain
\begin{equation}\label{5-6}
    \widetilde{V}(\widetilde{\alpha}_{i};z)-\widetilde{E}(\widetilde{\alpha}_{i})
    =f'^{2}\{V(\alpha_{i};f(z))-E(\alpha_{i})\}+\frac{1}{2}\{\frac{3}{2}(\frac{f''}{f'})^2-\frac{f'''}{f'}\}
\end{equation}
By substitution Eq.~(\ref{24}) into Eq.~(\ref{5-6}), we have
\begin{eqnarray}\label{5-7}
    \widetilde{V}(\widetilde{\alpha}_{i};z)-\widetilde{E}(\widetilde{\alpha}_{i})&=&f'^{2}(
    {-\frac{\alpha}{f}+\frac{\beta}{f^2}+\gamma-E})+\frac{1}{2}\{\frac{3}{2}(\frac{f''}{f'})^2-\frac{f'''}{f'}\}
\end{eqnarray}
We consider
\begin{equation}\label{5-8}
    f(z)=e^{-z}
\end{equation}
by this selection, one can define a point canonical transformation
as
\begin{eqnarray}\label{5-9}
    f(z)&=&e^{-z}\cr
    g(z)&=&e^{-\frac{z}{2}}
\end{eqnarray}
with above transformation, we can rewrite Eq.~(\ref{5-7}) as
\begin{eqnarray}\label{5-10}
    \widetilde{V}(\widetilde{\alpha}_{i};z)-\widetilde{E}(\widetilde{\alpha}_{i})
    =(\gamma-E+\frac{3}{4})\,e^{-2z}-\alpha\,e^{-z}+(\beta-\frac{1}{2})
\end{eqnarray}
which is like to Morse potential. In other words, by acting the
point canonical transformation Eq.~(\ref{5-9}) on the potential of
Eq.~(\ref{24}), that can explain the Kratzer and Coulomb potentials,
we obtain the Morse potential. In this situation we are looking for
this potential's algebra. For Morse potential
\begin{equation}\label{5-11}
    \widetilde{V}(z)=e^{-2z}-2be^{-z}
\end{equation}
the superpotential is
\begin{equation}\label{5-12}
    \widetilde{W}(z;a_{n})=a_{n}-e^{-z}
\end{equation}
Therefore the reminder in Eq.~(\ref{10}) is given by
\begin{equation}\label{5-13}
    R(a_{n})=2(a_{n}-1)
\end{equation}
where
\begin{equation}\label{5-14}
    a_{n}=b-(n+\frac{1}{2})
\end{equation}
One can use Eq.~(\ref{19})
\begin{equation}\label{5-15}
    R(a_{n})-R(a_{n-1})=-2
\end{equation}
Therefore, the commutation relation of Eq.~(\ref{24-2}) will vanish.
Now, we define the following dimensionless operators
\begin{equation}\label{5-16}
    \hat{K}_{0}:=\frac{1}{4}R(a_{0})
\end{equation}
and
\begin{equation}\label{5-17}
    \hat{K}_{\pm}:=\frac{1}{\sqrt{2}}\hat{B}_{\pm}
\end{equation}
where $\hat{B}_{\pm}$ has defined by Eq.~(\ref{16}). One can find
that the shape-invariant  algebra for these potentials is $SU(1,1)$
\begin{eqnarray}\label{5-18}
    [\hat{K}_{+},\hat{K}_{-}]&=&\frac{1}{2}[\hat{B}_{+},\hat{B}_{-}]\cr
    &=&2(-\frac{1}{4}R(a_{0}))\cr
    &=&-2\hat{K}_{0}
\end{eqnarray}
\begin{eqnarray}\label{5-19}
    [\hat{K}_{0},\hat{K}_{\pm}]&=&\frac{1}{4\sqrt{2}}[R(a_{0}),\hat{B}_{\pm}]\cr
    &=&\pm(\frac{4}{4\sqrt{2}}\hat{B}_{\pm})\cr
    &=&\pm\hat{K}_{\pm}
\end{eqnarray}

\section{Mapping of Hulth\'{e}n potential into P\"{o}schl-Teller potential 1}
The Hulth\'{e}n potential has the following form (we are using units
with $\hbar=1$, $2\,m=1$ )
\begin{equation}\label{6-1}
    V(r)=-\frac{e^{-r}}{1-e^{-r}}
\end{equation}
for mapping of this potential, we consider
\begin{equation}\label{6-2}
    f(z)=-2\ln{[\cos{z}]}
\end{equation}
by this selection, one can define a point canonical transformation
as
\begin{eqnarray}\label{6-3}
    f(z)&=&-2\ln{[\cos{z}]}\cr
    g(z)&=&\sqrt{-2\ln{[\cos{z}]}}
\end{eqnarray}
with above transformation, we can rewrite Eq.~(\ref{5-7}) as
\begin{eqnarray}\label{6-4}
    \widetilde{V}(\widetilde{\alpha}_{i};z)-\widetilde{E}(\widetilde{\alpha}_{i})
    =4(E-1)-\frac{1}{4}(1+16\,E)\sec^2{z}+\frac{3}{4}\csc^2{z}
\end{eqnarray}
which is like to P\"{o}schl-Teller potential 1. For
P\"{o}schl-Teller potential 1
\begin{equation}\label{6-5}
    \widetilde{V}(z)=-(A+B)^2+A(A-1)\sec^2{z}+B(B-1)\csc^2{z}
\end{equation}
the energy eignestates are given by
\begin{equation}\label{6-6}
    \widetilde{E}_{n}=(A+B+2\,n)^2-(A+B)^2
\end{equation}
Therefore the reminder in Eq.~(\ref{10}) is given by ( One can find
$R(a_{n})=\widetilde{E}_{n}-\widetilde{E}_{n-1}$)
\begin{equation}\label{6-7}
    R(a_{n})=4(2n+A+B-1)
\end{equation}
One can use Eq.~(\ref{19})
\begin{equation}\label{6-8}
    R(a_{n})-R(a_{n-1})=8
\end{equation}
Therefore, the commutation relation of Eq.~(\ref{24-2}) will vanish.
Now, we define the following dimensionless operators
\begin{equation}\label{6-9}
    \hat{K}_{0}:=\frac{-1}{8}R(a_{0})
\end{equation}
and
\begin{equation}\label{6-10}
    \hat{K}_{\pm}:=\frac{1}{2}\hat{B}_{\pm}
\end{equation}
where $\hat{B}_{\pm}$ has defined by Eq.~(\ref{16}). One can find
that the shape-invariant  algebra for these potentials is $SU(2)$
\begin{eqnarray}\label{6-11}
    [\hat{K}_{+},\hat{K}_{-}]&=&\frac{1}{4}[\hat{B}_{+},\hat{B}_{-}]\cr
    &=&2(-\frac{1}{8}R(a_{0}))\cr
    &=&2\hat{K}_{0}
\end{eqnarray}
\begin{eqnarray}\label{6-12}
    [\hat{K}_{0},\hat{K}_{\pm}]&=&\frac{-1}{16}[R(a_{0}),\hat{B}_{\pm}]\cr
    &=&\pm(\frac{1}{2}\hat{B}_{\pm})\cr
    &=&\pm\hat{K}_{\pm}
\end{eqnarray}

\section {conclusion}
For exactly solvable potentials of nonrelativistic quantum
mechanics, eigenvalues and eigenvectors can be derived using well
known methods of supersymmetric quantum mechanics. In this paper the
Schr\"{o}dinger equation with some potentials (Coulomb, Kratzer,
with Morse and P\"{o}schl-Teller type I with Hulth\'{e}n) has been
studied and we have shown that such potentials can be easily
inter-related among themselves within the framework of point
canonical coordinate transformations as the corresponding
eigenvalues may be written down in a closed form algebraically using
the well known results for the shape invariant potentials. Also we
have shown that the shape-invariant algebra for Coulomb, Kratzer,
and  Morse potentials is $SU(1,1)$, while the shape-invariant
algebra for P\"{o}schl-Teller type I and Hulth\'{e}n is $SU(2)$. We
must mention that the Morse potential is also related with the SU(2)
group except for SU(1,1) one, to see the SU(2) group approach refere
to \cite{dong1}.

\end{document}